\documentclass[12pt]{article}
\usepackage{psfig,epsfig}
\def\Frac#1#2{\frac{\displaystyle{#1}}{\displaystyle{#2}}}

\def\21{$SU(2) \ot U(1)$}
\def\321{$SU(3) \ot SU(2) \ot U(1)$}

{\bf

\def\n.c.#1#2#3{         { Nuovo Cim. }{\bf #1}, #3 (19#2)}
\def\r.n.c.#1#2#3{       { Riv. del Nuovo Cim. }{\bf #1}, #3 (19#2)}

\begin{document}
\thispagestyle{empty}
\begin{titlepage}
\begin{center}
\hfill hep-ph/yymmdd\\
\hfill FTUV/99-38\\
\hfill IFIC/99-39\\
%\hfill{version of \today}\\
\vskip 0.5cm
{\bf \large
Probing Neutrino Magnetic Moments at Underground Detectors
with Artificial Neutrino Sources
}
\end{center}
\normalsize
%\vskip1cm
\begin{center}
{\bf O. G. Miranda~$^a$}\footnote{ E-mail: omr@fis.cinvestav.mx}
{\bf J. Segura~$^b$\footnote{ E-mail: segura@flamenco.ific.uv.es}},
{\bf V. B. Semikoz~$^c$},\\
\vskip .3cm
{\bf and  J.~W.~F. Valle~$^d$}\footnote{ E-mail valle@flamenco.ific.uv.es}\\
\end{center}
%
%
%\begin{center}
%\baselineskip=13pt
%{\it $^a$ Institute for Nuclear Research, Moscow, Russia}
%\vglue 0.8cm
%\end{center}

\begin{center}
{\it $^a$  Departamento de F\'{\i}sica \\ CINVESTAV-IPN, A. P. 
14-740, M\'exico 07000, D. F., M\'exico.} \\
\end{center}
\begin{center}
{\it $^b$  Instituto de Bioingenier\'{\i}a\\ 
Universidad Miguel Hern\'andez \\
Edificio La Galia\\
03202 Elche (Alicante) SPAIN} \\
%\vglue 0.8cm
\end{center}
\begin{center}
{\it $^c$ The Institute of the Terrestrial Magnetism, \\ 
the Ionosphere and Radio Wave Propagation of the Russian Academy of Science, \\
IZMIRAN, Troitsk, Moscow region, 142092, Russia} \\
%\vglue 0.8cm
\end{center}
\begin{center}
{\it $^d$ Instituto de F\'{\i}sica Corpuscular - C.S.I.C.\\
Departament de F\'{\i}sica Te\`orica, Universitat de Val\`encia\\}
{\it 46100 Burjassot, Val\`encia, SPAIN \\
http://neutrinos.uv.es        }\\
%\vglue 0.8cm
\end{center}
\baselineskip=13pt

\begin{abstract}

Neutrino-electron scattering can be used to probe neutrino
electromagnetic properties at low-threshold underground detectors with
good angular and recoil electron energy resolution. We propose to do
this using a number of artificial neutrino and anti-neutrino sources
such as $^{51}Cr_{24}$ and $^{90}Sr-Y$. The neutrino flux is known to
within one percent, in contrast to the reactor case and one can reach
lower neutrino energies.  For the $^{90}Sr-Y$ source we estimate that
the signal expected for a neutrino magnetic moment of $\mu_{\nu}=6
\times 10^{-11}\mu_{B}$ will be comparable to that expected in the SM
and corresponds to a 30\% enhancement in the total number of expected
events.  \\ pacs{13.15.+g 12.20.Fv 14.60.St 95.55.Vj }
\end{abstract}
%pacs{13.15.+g 12.20.Fv 14.60.St 95.55.Vj }
\end{titlepage}
%\pacs{13.15.+g 12.20.Fv 14.60.St 95.55.Vj }
\vskip 1cm

\section{Introduction}

Low-energy-threshold underground detectors with good angular and
recoil electron energy resolution open a new window to probe the
structure of the weak interaction and neutrino electromagnetic
properties \cite{Vogel} \cite{Barabanov:1998bj} as a good alternative
to what can be learned at reactor and accelerator
experiments~\cite{Vogel,Moretti:1998py}.  We propose to do this using
a number of artificial neutrino and anti-neutrino sources such as
$^{51}Cr_{24}$ and $^{90}Sr-Y$. The neutrino flux is known to within a
one percent accuracy, in contrast to the reactor case and one can
reach lower neutrino energies.

Non-standard neutrino properties have been studied for several years,
partly motivated by the solar neutrino problem. A possible explanation
of this problem is related to a large neutrino magnetic moment
\cite{Akhmedov}. Present constraints for the electron neutrino
magnetic moment coming from reactor experiments gives
$\mu_{\nu}=1.8\times 10^{-10}\mu_{B}$ \cite{Derbin}. The improvement
of this bound in a new reactor experiment is the goal of the MUNU
\cite{MUNU} experiment, now running.

The idea of using an artificial neutrino source (ANS) to search for a
neutrino magnetic moment was first put forward by Vogel and Engel in
Ref. \cite{Vogel}.  This kind of sources have already been used to
calibrate both GALLEX and SAGE experiments \cite{GALLEX} and,
recently, this idea has been considered by several experimental groups
working in underground physics.  The ANS have as an advantage that the
uncertainties in the neutrino flux intensity are lower than in the
case of reactor neutrinos and they have a small size, which makes them
suitable for a deep underground experiment; they could be even
surrounded by the detector as is the plan for the LAMA collaboration
\cite{LAMA} that has as a goal the use of a $^{147}~Pm$ anti-neutrino
source with a one ton $NaI$ detector in order to test for a neutrino
magnetic moment in the region $10^{-11}\mu_{B} <\mu_{\nu}
<10^{-10}\mu_{B}$. The BOREXINO collaboration has also the possibility
of searching for a neutrino magnetic moment in such a region using an
ANS located at a distance of the order of 10 m \cite{Ferrari,Ianni}.

Here analyse the potential of these sources in testing the neutrino
magnetic moment in a detector with both angular and recoil electron
energy resolution. Such a study could be interesting for a detector
like that in the HELLAZ proposal \cite{HELLAZ} that is planning to
detect neutrinos through neutrino electron scattering with an energy
threshold as low as 100~KeV and with and angular resolution of 35
mrad. These two characteristics could make HELLAZ proposal adequate
for improving the limits on neutrino properties by using artificial
neutrino sources. The only limitation HELLAZ could have in comparison
with BOREXINO is the large mass the last experiment is planning (100
tones vs. 6 tones) although this might be solve with an adequate
experimental set up; for example, if it were possible to surround the
source with the detector to get the full $4\pi$ neutrino source of the
source, although in this case oscillation could not be studied.

\section{Artificial Sources and Magnetic Moment}

We will consider the next sources for being
the most realistic and interesting ones in getting results:

\begin{figure}
\centerline{\protect\hbox{\psfig
{file=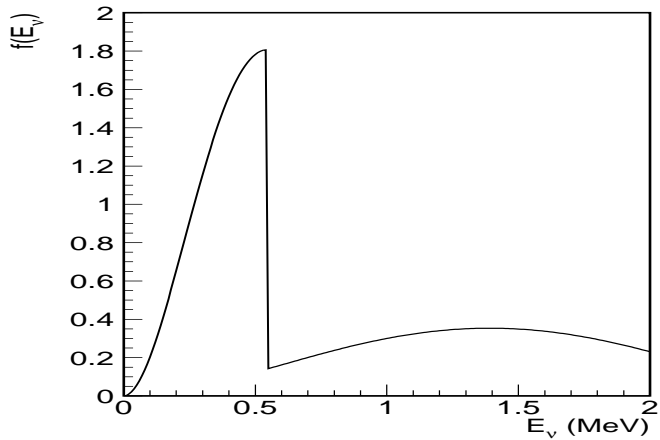,height=6cm,width=10cm,angle=0}}}
\caption{Sr-Y spectrum }
\end{figure}

\begin{itemize}
%%\begin{enumerate}
\item $^{51}Cr_{24}$ neutrino source. This is a neutrino source
that has already been used for calibrating both SAGE and GALLEX
experiments \cite{GALLEX}. The main neutrino line is $E_{\nu}=746$ KeV
and the lifetime is 40 days.

\item $^{49}V_{23}$ This is a neutrino source that produces neutrinos 
with energy $E_{\nu}=602$ KeV. The lifetime is 1.3 years \cite{tom}.

\item $^{145}Sm_{62}$ This is a neutrino source that produces neutrinos 
with energy $E_{\nu}=554$ KeV. The lifetime is 1.34 years \cite{tom}.

\item $^{37}Ar_{18}$ This is a neutrino source that produces neutrinos 
with energy $E_{\nu}=814$ KeV. The lifetime is 35 days \cite{tom}.  

\item  $^{90}Sr$ anti-neutrino source. This source has been
studied in Ref. \cite{Bergelson} and its potential for the BOREXINO
case has been already discussed \cite{Ianni}.  The neutrino energy
spectrum for such a source is shown in Fig 1.  The half-life is 28
years.

%%\end{enumerate}
\end{itemize}

For the case of neutrino sources the detectors which are now being
proposed would be able to measure the differential cross section for
the $\nu_{e} e$ scattering. At leading order in the SM this is given
by
\begin{eqnarray}
\frac{d\sigma^{W}}{dT}  = 
 \frac{2m_{e}G^2_{F}}{\pi} \big\{ 
   g_{L}^2+g_{R}^2 (1-\frac{T}{E_{\nu}})^2  - 
   \frac{m_e}{E_{\nu}} g_{R}
    g_{L}\frac{T}{E_{\nu}} \big\}  \label{DCS}
\end{eqnarray}
where $T$ is the recoil electron energy, and $E_{\nu}$ is the neutrino
energy; $g_{L}=\frac12+sin^2\theta_{W}$ and $g_{R}=sin^2\theta_{W}$.  

For the case of an anti-neutrino source the process will be
$\bar{\nu_{e}}-e$ scattering and we just need to exchange $g_{L}$ with
$g_{R}$ in order to get the corresponding differential cross section.

On the other hand the differential cross section in the case of a
neutrino magnetic moment is 

\begin{eqnarray}
\frac{d\sigma^{mm}}{dT}  =
 \frac{\pi\alpha^2\mu^2_{\nu}}{m^2_{e}}\big\{  \frac1T -\frac{1}{E_{\nu}}
\big\}  \label{MMCS}
\end{eqnarray}

\noindent
which adds incoherently to the weak cross section (neglecting neutrino
mass).

It is well known from this equation that for lower values of T the
neutrino magnetic moment signal will drastically increase. This makes
interesting the use of low threshold detectors such as the one has
been proposed by the HELLAZ collaboration, and which could reach a
threshold as low as 100 KeV.

Besides the low energy threshold, HELLAZ will also have angular
resolution. This could be useful not only to lower the systematic
errors, but also to take advantage of the best regions in the
($\theta$,$T$) plane where the non-standard effects could be
bigger. Although the restriction to a narrow window will limit the
statistics, the enhancement of the neutrino magnetic moment (NMM)
effect may over-compensate and one might have an overall gain.

In \cite{Segura2} it was shown that $\frac{d\sigma^{W}}{dT}$ vanishes
for forward electrons (which implies maximum recoil energy for $e^{-}$)
for a $\bar{\nu}_{e}$ energy  given by:

\begin{equation}
E_{\nu}=m_{e} \Frac{g_{L}-g_{R}}{2g_{R}}=m_{e}/4sin^2\theta_{W}\simeq
0.548 MeV
\end{equation}

This kind of cancellation only takes place when considering scattering
of $\bar{\nu}_{e}$ off $e^{-}$ Of course, to be able to study this
effect, experiments capable of measuring both the recoil (kinetic)
energy of the electron (T) and its recoil angle ($\theta$) become
necessary, so that we can select neutrino energies $E_{\nu}$ from a
non-monochromatic source (a monochromatic source of $
\bar{\nu}_{e}$ with $E_{\nu}=m_{e}/4sin^2\theta_{W}$ would be the ideal but 
there are not monochromatic anti-neutrino sources).

The three variables $E_{\nu}$, $T$ and $\theta$ are related by the
equation:

\begin{equation}
cos\theta = \frac{T}{\sqrt{T^2+2m_{e}T}}\big( 1+\frac{m_{e}}{E_{\nu}}  \big)
\label{cos}
\end{equation}

As discussed in \cite{Segura2} the dynamical zero seems potentially
interesting to measure $\mu_{\nu}$ since it opens a window in phase
space where the weak cross section becomes small, so that the magnetic
moment contribution could eventually become larger than the weak cross
section.  However, as discussed in \cite{javi} in the context of
reactor neutrino experiments, the fact that the statistics close to
the dynamical zero is poor is, unfortunately, more important than the
enhancement in $d\sigma^{mm}/d\sigma^{W}$.

\subsection{Neutrino sources }

\begin{figure}
\centerline{\protect\hbox{\psfig{file=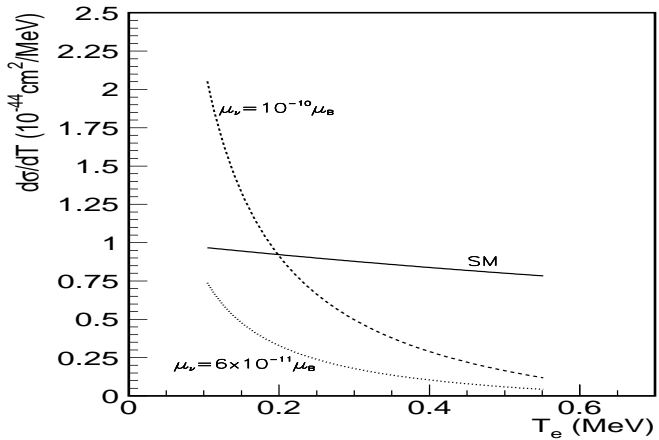,height=6cm,width=10cm,angle=0}}}
\caption{Differential cross section in terms of $T$ for the SM case
and for the neutrino magnetic moment contribution for two different values
of $\mu_{\nu}$ in the case of a Cr
source.}
\label{dcsTdr}
\end{figure}

We will consider first the case of a $Cr$ neutrino source.  As
mentioned in the introduction this source has already been used for
the calibration of SAGE and GALLEX experiments. We will consider the
746 KeV neutrino line that has the 81 \% of the neutrino flux. 

For this case the differential cross section will be as given in Eq.
(\ref{DCS}) and we just need to substitute the corresponding value for
the neutrino energy. The result is shown in Fig 2 both for the
Standard Model case as well as for the case of a neutrino magnetic
moment. Different values of the magnetic moment are shown.   As we
can see, for low values of $T$ the NMM signal is of the same order of
the SM one for $\mu_{\nu}\sim .6-1\times 10^{-10}\mu_{B}$.

The differential cross section
\begin{equation}
\frac{d\sigma^{W}}{dcos\theta} =\frac{d\sigma^{W}}{dT}\frac{dT}{dcos\theta}
\end{equation}
can be easily obtained and it is shown in Fig. 3, where we can see a
similar result: there is a region, for large electron recoil angle,
for which the NMM signal is comparable to that of the SM. The
similarities of these two figures are more evident if one notices that
the recoil angle $\theta$ is maximum for lower T, as can be derived
from Eq. (\ref{cos}).

\begin{figure}
\centerline{\protect\hbox{\psfig{file=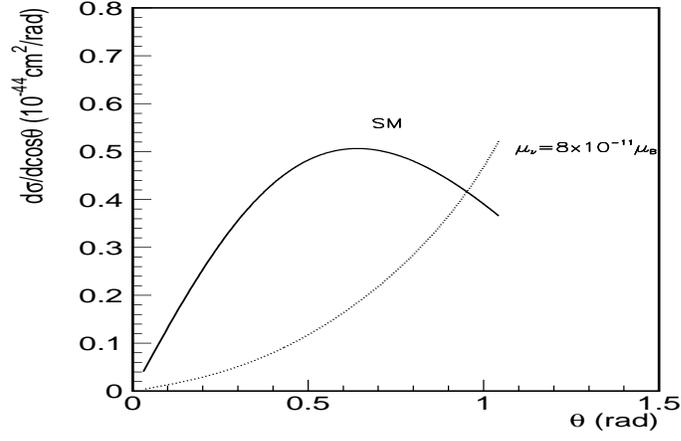,height=6cm,width=10cm,angle=0}}}
\caption{Differential cross section in terms of cos$\theta$ for the SM case 
and for the case of a neutrino magnetic moment 
$\mu_{\nu}=8\times 10^{-11}$ in the case of a Cr source. 
%{\it 
% Remark 1: the notation
% $\zeta$ ($\equiv \theta$ ) should be changed in the figure; Remark 2: 
% figures 2 and 3 
% are related through the factor $dT/d(\cos\theta)$}
}
\label{dcscdr}
\end{figure}

For other neutrino sources different that the $Cr$ source, the shape
of the plots shown in Fig. 2 and Fig. 3 will have a slightly change,
due to the fact that the neutrino energy is different. However the
qualitatively result will be the same. 

\subsection{Anti-neutrino sources }

Now we consider the case of a $^{90}Sr-^{90}Y$ anti-neutrino
source. This source has been studied by a Moscow group
\cite{Bergelson} and its potential has been studied for the BOREXINO
case \cite{Ianni}.

\begin{figure}
\centerline{\protect\hbox{\psfig{file=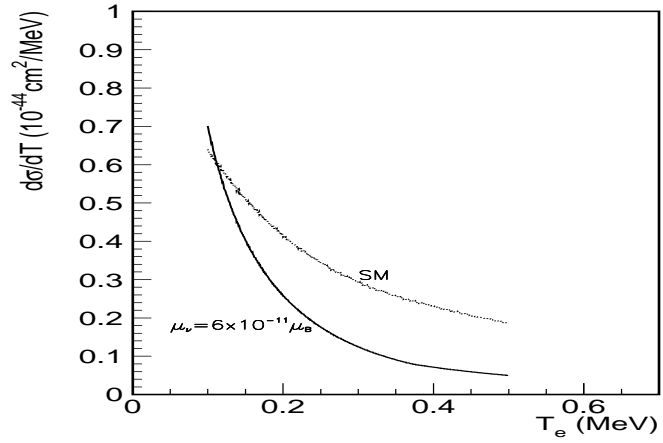,height=6cm,width=10cm,angle=0}}}
\caption{T-distribution of events for the SM case  and for the case 
of a neutrino magnetic moment $\mu_{\nu}=6\times 10^{-11}$ for a
$Sr-Y$ source. The angle $\theta$ has been integrated over.}
\label{tsr-y}
\end{figure}

We can make a similar analysis as in the case of the neutrino sources
in the previous section. The main difference here, beside the
interchange of $g_{L}$ by $g_{R}$ in the differential cross section,
is that we have an energy spectrum instead of a neutrino energy line.
Therefore, we need integrate over the neutrino energy distribution. As
we are interested in the angular distribution, it is more convenient
to use Eq. (\ref{cos}) to make a change of variables in the
integration.  This has also been done in
\cite{javi}. The result can be express as

\begin{eqnarray}
{\langle \frac{d\sigma}{dT} \rangle}_{E_\nu}
=\int{\frac{d^2 \sigma}{dTd(cos\theta )}d(cos\theta )}  
= \int{\Theta_{\mbox{p. s.}} f(T,\theta ) 
             \frac{d \sigma(T,\theta )}{dT} 
	     \frac{m_{e}pT}{(pcos\theta -T)^2}d(cos\theta )} \label{dif}
\end{eqnarray}
or, if we are interested in the angular distribution 

\begin{eqnarray}
{\langle \frac{d\sigma}{d(cos\theta )}\rangle}_{E_\nu}
=
\int{\frac{d^2 \sigma}{dTd(cos\theta )}dT}  
= \int{\Theta_{\mbox{p. s.}} f(T,\theta ) 
             \frac{d \sigma(T,\theta )}{dT} 
	     \frac{m_{e}pT}{(pcos\theta -T)^2}dT} \label{dif2}
\end{eqnarray}

%\begin{eqnarray}
%\frac{d\sigma}{dT}=\int{\frac{d^2 \sigma}{dTd(cos\theta )}d(cos\theta )}  
%= \int{\Theta_{\mbox{p. s.}} f(T,\theta ) 
%             \frac{d \sigma(T,\theta )}{dT} 
%	     \frac{m_{e}pT}{(pcos\theta -T)^2}d(cos\theta )} \label{dif}
%\end{eqnarray}
%
%or, if we are interested in the angular distribution 
%
%\begin{eqnarray}
%\frac{d\sigma}{d(cos\theta )}=
%\int{\frac{d^2 \sigma}{dTd(cos\theta )}dT}  
%= \int{\Theta_{\mbox{p. s.}} f(T,\theta ) 
%             \frac{d \sigma(T,\theta )}{dT} 
%	     \frac{m_{e}pT}{(pcos\theta -T)^2}dT} \label{dif2}
%\end{eqnarray}

\noindent
In this equations $\Theta_{\mbox{p. s.}}$ accounts for the allowed
phase space and $f(T,\theta)=f(E_{\nu}(T,\theta))\equiv dn/d E_{\nu}$
is the neutrino energy spectrum as a function of $T$ and $\theta$.

\begin{figure}
\centerline{\protect\hbox{\psfig{file=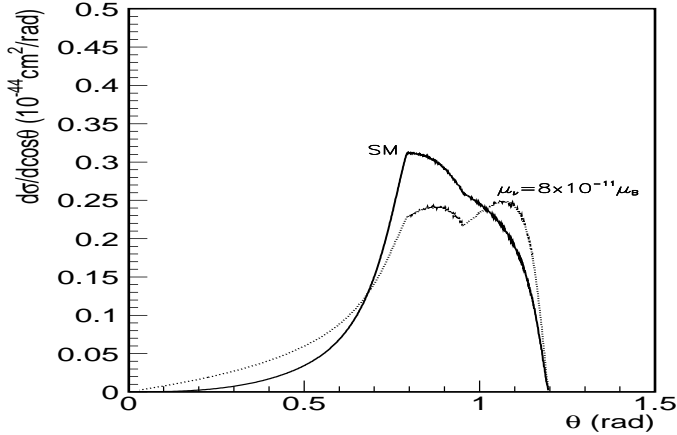,height=6cm,width=10cm,angle=0}}}
\caption{Angular distribution of events for the SM case 
and for the case of a neutrino magnetic moment 
$\mu_{\nu}=6\times 10^{-11}$. We consider a $Sr-Y$ source (Fig.1) and a 
threshold of $T_{th}=100KeV$; the recoil energy is integrated, considering the
cuts imposed by the threshold and the kinematical limits. 
%{\it Remark: again, $\zeta$ should be changed by $\theta$ (or $\theta$ 
%by $\zeta$ in 
%the text)}
}
\label{sr-y}
\end{figure}

We have computed the differential cross sections $d\sigma /dT$ and
$d\sigma /dcos\theta$ for an anti-neutrino energy spectrum given as
shown in Fig. 1, normalized to one. In the case of $d\sigma /dT$ we
have integrated Eq.  (\ref{dif}) in the whole allowed $\theta$ range.
For the case of the differential cross section $d\sigma /dcos\theta$
we have integrated $T$ in the range $.1 MeV< T < 0.5 MeV$, the energy
range to which HELLAZ could be sensitive.  The results are shown in
Fig. \ref{tsr-y} and Fig \ref{sr-y} both for the Standard Model case
as well as for the case of a neutrino magnetic moment. Different
values of the magnetic moment are shown in these figures.  The kink in
the distribution in Fig. \ref{sr-y} is due to the sharp decrease in
the energy spectrum for $E_{\nu}>0.5 MeV$ (fig. 1).

Notice that, besides there is an additional contribution to the
differential cross section, the shape is also different from that of
the standard model. In particular the magnetic moment contribution is
slightly bigger than the Standard Model one both for small angles and
for big angles, meanwhile, in the intermediate region, the SM is
bigger. However, the low values of the differential cross section in
the small angle region, in comparison with other angles could make the
analysis of this region more difficult. The case of the large angle
region can be easily explained if we consider that, for minimum recoil
electron energy the recoil angle is maximum, therefore, as the
magnetic moment contribution increases at low $T$, it is natural to
have a similar effect for large $\theta$.

\begin{figure}
\centerline{\protect\hbox{\psfig{file=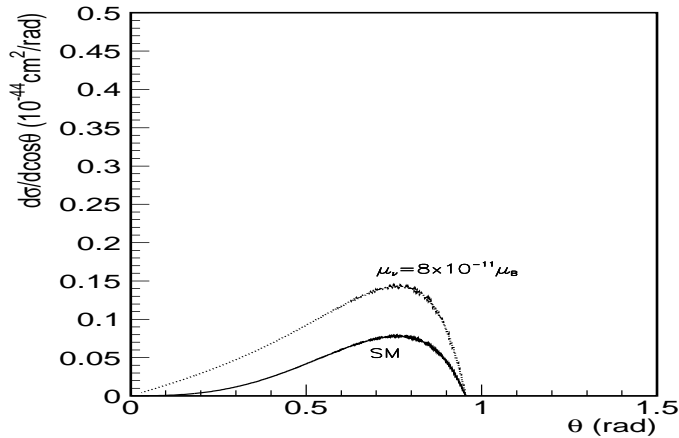,height=6cm,width=10cm,angle=0}}}
\caption{Same as in the previous figure  but considering a 
hypothetical $Sr$ source consisting only in the low energy part of the
spectrum in Fig. 1 (neglecting the Y spectrum).  Notice that the cross
sections are smaller since we have cut part of the spectrum; the
magnetic moment contribution becomes larger that the weak term since
we restrict the spectrum to lower energies.}
\label{sr}
\end{figure}

Besides the advantages that ANS have in general, this particular
source seems to be interesting because the energy range of the
spectrum that belongs to the $^{90}Sr$ has a peak in an energy range
that is close to the kinematical zero that has been already discussed
in the previous section. In order to illustrate the potential of this
energy region we show in Fig. \ref{sr} the result that would be
obtained for the ideal hypothetical case of a pure $^{90}Sr$ without
any contaminant.

\section{The NMM Signal and Recoil Angle Resolution}

We now come back to the real $Sr-Y$ source. In this case one could try
to optimise the best region in the $\theta -T$ plane on which the
non-standard effect is maximum. In order to do this analysis let us
consider the curves in the $(T,\theta)$ plane given by the condition

\begin{equation}
C=\Frac{d\sigma^{mm}/dT}{d\sigma^{W}/dT}
\end{equation}

\begin{figure}
\centerline{\protect\hbox{\psfig{file=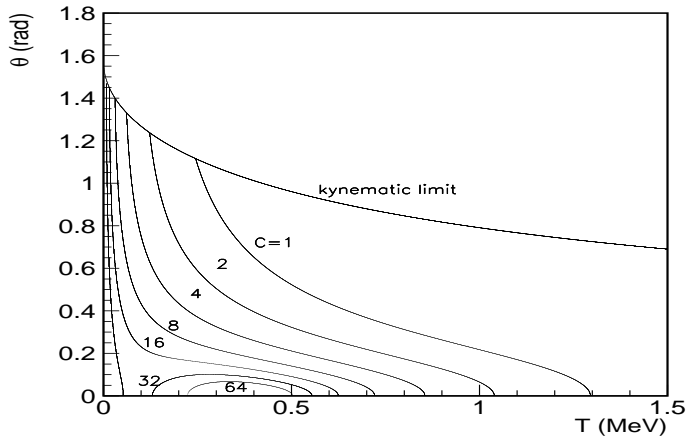,height=6cm,width=10cm,angle=0}}}
\caption{
Curves of equal ratio $C\equiv 
\frac{d\sigma^{mm}}{dT}/\frac{d\sigma^{W}}{dT}$ for
$\bar{\nu}_{e}$ and taking $\mu_{\nu}=10^{-10}\mu_{B}$. Two are the effects
which increase the ratio $C$: for low $T$ the magnetic moment contribution
 becomes larger; for values of $T$ and $\theta$ near the dynamical zero the
weak term tends to cancel.
}
\label{ratiosd}
\end{figure}

This gives us the curves shown in Fig. 7. These are characterized by a
given ratio of the magnetic moment differential cross section to the
SM one. Therefore, for C=1 we will get the curve where the magnetic
moment signal is equal than the SM one, for C=2 the the magnetic
moment signal is twice the SM one, and so on.
The corresponding iso-curves are given in figure \ref{ratiosd}, for
$\mu_{\nu}= 10^{-10}\mu_{B}$. In the figures the curves for
$c=1,2,4,8,32$ are shown; of course, for different selection of
magnetic moments, the values of $c$ in the same figure are scaled. For
instance, taking $\mu_{\nu}=10^{-11}\mu_{B}$, the curves shown in
Fig. 4 would correspond to $c=0.01,0.02,0.04,0.08,0.32$.  It is also
important to notice that, for any couple ($\theta$,$T$) the neutrino
energy is already fixed by the kinematics as can be seen from
Eq. (\ref{cos}). 

\begin{figure}
\centerline{\protect\hbox{\psfig{file=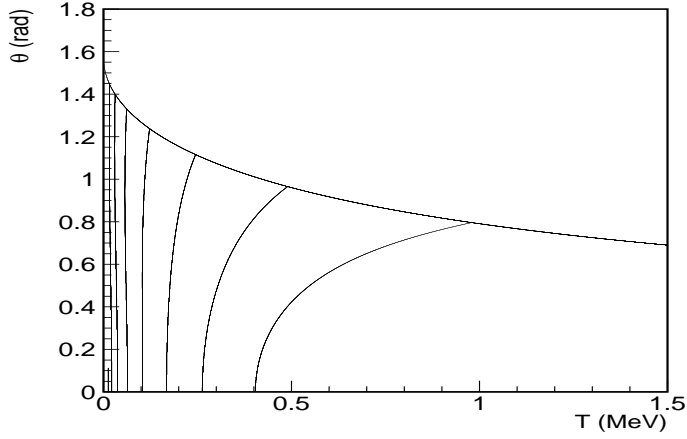,height=6cm,width=10cm,angle=0}}}
\caption{
Curves of equal ratio $d\sigma^{mm}/d\sigma^{W}$ for
$\nu_{e}$. $\mu_{\nu}=10^{-10}\mu_{B}$
}
\label{niso}
\end{figure}

The effect of the dynamical zero on the iso-curves can be noticed
specially in the cases of ratios $c=32,64$, where curves surrounding
the position of the dynamical zero appear. This effect does not appear
in the case of a neutrino source as can be seen in Fig. \ref{niso}
where similar curves are shown for the case of $\nu_e - e$ scattering.

Given that the iso-curves (Fig. \ref{ratiosd}) reflect the presence of
a favoured region for searching for a magnetic moment, thanks to the
dynamical zero, it seems interesting to integrate the cross section
over regions in the $(T,\theta)$ plane limited by the iso-curves. In
this way, we are optimising the region of integration to look for
magnetic moment.

Figure 9 shows the result of integrating the differential cross
section given in Eq. (\ref{dif}) over $T$ and $\theta$ in regions such
that $d\sigma^{mm}/d\sigma^{W} >C$.  A neutrino magnetic moment
$\mu_{\nu}=6\times 10^{-11} \mu_{B}$ has been assumed.  Of course, as the
limiting ratio $C$ is taken larger, the magnetic moment signal becomes
larger than the SM one.  However, the integral in this case is small.
Note that if one integrates over the whole region (C=0) one can probe
the complete cross section, but the value of NMM relative to SM
decreases. Therefore it is interesting to study intermediate regions
such as the region limited by $C=0.7$, where the contribution of the
neutrino magnetic moment is comparable with that of the weak
interaction, although the statistics is a 30 \% of the total one.

\begin{figure}
\centerline{\protect\hbox{\psfig{file=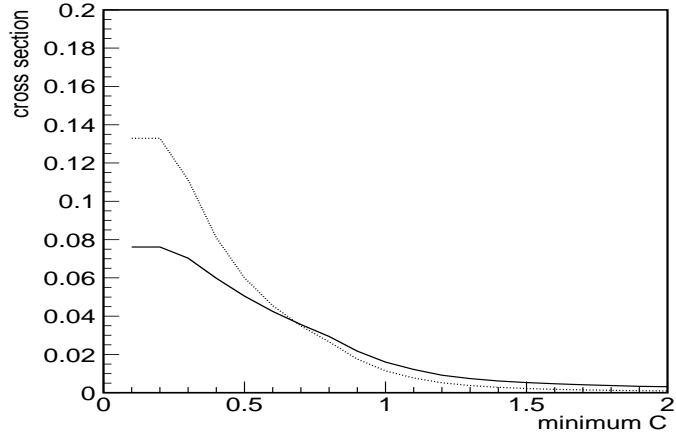,height=6cm,width=10cm,angle=0}}}
\caption{Weak and magnetic moment ($\mu_{\nu}=6\times 10^{-11}\mu_{B}$) 
integrated cross sections; we integrate both over angles $\theta$ and
energies $T$. The region of integration is limited by 
the curves of equal ratio $C$ ; the limiting
 value $C$ is
displayed in the horizontal axis; also, we consider a threshold for $T$:
$T\ge T_{th}=100 \,KeV$. Both  variables ($T$,$\theta$)
are also limited by the kinematics.} 
\label{ratio}
\end{figure}

\section{Discussion \& Conclusions}

We have discussed the potential of investigating neutrino-electron
scattering as a probe of neutrino magnetic moment of the order
$\mu_{\nu} \sim 10^{-11}\mu_{B}$ at low-threshold underground
detectors with good angular and recoil electron energy resolution.  We
propose to do this using a number of artificial neutrino and
anti-neutrino sources such as $^{51}Cr_{24}$ and $^{90}Sr-Y$.  The
neutrino flux is known to within a one percent accuracy, in contrast
to the reactor case and one can reach lower neutrino energies.  For
the $^{90}Sr-Y$ source we have investigated the possible role of
dynamical zeros in improving the sensitivity to the neutrino magnetic
moment, with a negative result due to the poor statistics in this
region. However, integrating over large kinematical regions, we
estimated that the signal expected for a neutrino magnetic moment of
$\mu_{\nu} = 6 \times 10^{-11}\mu_{B}$ will be comparable to that
expected in the Standard Model and corresponds to a 30\% enhancement
in the total number of expected events.  In order provide a more
reliable estimate of the sensitivities to the neutrino magnetic moment
that can be reached in this kind of studies a dedicated experimental
analysis will be necessary.

\section*{Acknowledgements}

This work was supported by DGICYT grant PB95-1077, by Intas Project
96-0659 and by the EEC under the TMR contract ERBFMRX-CT96-0090. OGM
was supported by SNI-M\'exico and by the grant CINVESTAV 
JIRA '99/08. VBS was supported by the RFFR grant 97-02-16501 and by
Generalitat valenciana. We thank Tom Ypsilantis for discussions.


\newpage

\begin{thebibliography}{9}
\bibitem{Vogel} 
P. Vogel and J. Engel, Phys. Rev. {\bf D39} 3378 (1989).

\bibitem{Barabanov:1998bj}
I.~Barabanov {\it et al.}, Nucl. Phys. {\bf B546} 19-32 (1999), hep-ph/9808297;
O.G.~Miranda, V.~Semikoz and J.~W.~F. Valle, Phys. Rev. {\bf D58}, 013007
(1998), hep-ph/9712215; O. G.~Miranda, V.~Semikoz 
and J.~W.~F.~Valle, ``Neutrino electron scattering as a probe of the
electroweak gauge structure," {\it In Valencia 1997, Beyond the
standard model 340-344}, Ed.  I. Antoniadis, L. Ibanez,
J.~W.~F.~Valle, World Scientific Publishing Co., 1998, ISBN
981-02-3638-7.

\bibitem{Moretti:1998py}
M.~Moretti, C.~Broggini and G.~Fiorentini,
``Physics beyond the standard model with a new reactor experiment,"
Phys. Rev. {\bf D57}, 4160 (1998).

\bibitem{Akhmedov} E. Kh. Akhmedov hep-ph/9705451.

\bibitem{Derbin} Cernyi et. al., Phys. of At. Nuc. {\bf 57}
222 (1994).

\bibitem{MUNU}
MUNU Coll. C. Amsler et. al., Nucl. Inst. and Meth. A369 115 (1997). 

\bibitem{GALLEX} 
GALLEX Coll., {\it Phys. Lett.} {\bf B342} 440 (1995); SAGE Coll.,
J. N. Abdurashitov, {\it et. al.} {\it Phys. Rav. Lett.} {\bf 23}
4708 (1996).

\bibitem{LAMA} I. R. Barabanov et. al. Astro. Phys. {\bf 8} 67 (1997).

\bibitem{Ferrari} N. Ferrari, C. Fiorentini, B. Ricci, Phys. Lett. {\bf B387} 
 427 (1996).

\bibitem{Ianni} A. Ianni, D. Montanino, G. Scioscia hep-ex/9901012.

\bibitem{HELLAZ} F. Arzarello, CERN-LAA 94-19.

\bibitem{tom} T. Ypsilantis, private com.

\bibitem{Segura2} J. Segura et al., Phys. Rev. D49 1633 (1994).

\bibitem{javi}
J. Segura, Europhys. Jour. C  (1998)

\bibitem{Bergelson} B. R. Bergelson, A. V. Davydov, Yu. N. Isaev, and
V. N. Kornoukhov, Phys. of At. Nuc. {\bf 61} 1347 (1998).

\end{thebibliography}
\end{document}